\documentclass[journal,twoside]{IEEEtran}
\usepackage{orcidlink}
\usepackage{cite}
\usepackage{amsmath,amssymb,amsfonts}
\usepackage{algorithmic}
\usepackage[braket, qm]{qcircuit}
\usepackage{graphicx}
\usepackage{subcaption}
\usepackage{textcomp}
\usepackage{xcolor}
\usepackage{comment}
\usepackage{hyperref}

\title{Digital Quantum Reservoir Computing for ATM Time Series Prediction
}

\author{\IEEEauthorblockN{Chiara Vercellino\IEEEauthorrefmark{1}\orcidlink{0000-0002-0562-3157},
Giacomo Vitali\IEEEauthorrefmark{1}\orcidlink{0000-0002-3056-796X},
Valeria Zaffaroni\IEEEauthorrefmark{3}\orcidlink{0009-0002-8221-5904},
Francesca Cibrario\IEEEauthorrefmark{3}\orcidlink{0009-0007-8290-4992},\\
Emanuele Dri\IEEEauthorrefmark{1}\orcidlink{0000-0002-5144-1514},
Paolo Viviani\IEEEauthorrefmark{1}\orcidlink{0000-0001-8947-9481},
Olivier Terzo\IEEEauthorrefmark{1}\orcidlink{0000-0001-8482-2607},
and Davide Corbelletto\IEEEauthorrefmark{3}\orcidlink{0009-0003-8830-2619}}\\
\IEEEauthorrefmark{1}\textit{LINKS Foundation}, Torino, Italy - 
    chiara.vercellino@linksfoundation.com, giacomo.vitali@linksfoundation.com, \\
    emanuele.dri@linksfoundation.com,
    paolo.viviani@linksfoundation.com,
    olivier.terzo@linksfoundation.com, \\
\IEEEauthorrefmark{3}\textit{Intesa Sanpaolo}, Torino, Italy -
valeria.zaffaroni@intesasanpaolo.com,  francesca.cibrario@intesasanpaolo.com, \\ davide.corbelletto@intesasanpaolo.com}
\begin{document}
\maketitle

\begin{abstract}
We investigate a digital quantum reservoir computing (QRC) framework for multi-step forecasting of automated teller machine (ATM) cash demand time series on near-term quantum devices. The proposed approach uses parametrized four-qubit reservoirs with a fixed structure exploiting partial measurement and reset, where temporal data is encoded in rotation angles. Training is restricted to a classical Ridge-regression readout. We systematically analyze the impact of the circuit ansatz\"e, reservoir memory, measurement-derived observables, and the execution backend on the forecasting performance. Experiments are performed with noiseless simulation, noise-aware emulation, and a real IQM Spark quantum processor. Although the QRC models do not outperform the classical Prophet benchmark in terms of Mean Absolute Error and Normalized Mean Squared Error metrics, they achieve more competitive results in Dynamic Time Warping metric, indicating a partial ability to capture temporal structure. These findings provide an empirical assessment of digital QRC for realistic financial forecasting and highlight both its current limitations and its potential on near-term quantum hardware.
\end{abstract}


\section{Introduction}

Dynamical quantum systems are increasingly investigated as computational substrates for processing temporal data. Within this paradigm, quantum reservoir computing (QRC) offers a particularly attractive approach: a fixed quantum system provides a high-dimensional nonlinear feature map, while training is restricted to a classical readout layer. By avoiding gradient-based optimization of deep parameterized circuits, QRC mitigates trainability issues and reduces circuit design overhead, making it well suited to near-term noisy intermediate-scale quantum (NISQ) devices, where coherence time, gate fidelity, and circuit depth are limiting resources.

Digital implementations of QRC have demonstrated that gate-based quantum circuits can emulate reservoir dynamics with fine-grained control over entanglement structure, measurement strategy, and input encoding. In this setting, computational expressivity arises primarily from architectural and protocol choices rather than parameter optimization. Circuit ansatz, measurement back-action, observable selection, and circuit depth collectively determine the effective memory, nonlinearity, and feature space of the reservoir. Therefore, understanding how these design elements shape the temporal information processing is a central question for digital QRC.

In this work, we investigate a digital quantum reservoir architecture and systematically explore how circuit design and measurement choices impact time series prediction performance. Our workflow processes a univariate time series in a sequential manner. At each time step preceding the one to be predicted, the scalar input value is encoded as a rotation angle in a quantum circuit ansatz. Measurements on qubits produce classical observables that form the feature vector for a linear model whose task is to predict future values of the series.

We consider architectures based on partial measurement of the quantum system. The reservoir consists of four qubits: two \textit{auxiliary} qubits that are measured and reset at every time step, and two \textit{system} qubits that preserve information from previous time steps, thereby providing temporal memory. We study two ansätze: a baseline ansatz~\cite{yasuda2023quantum, chen2020temporal} and a Multiscale Entanglement Renormalization Ansatz (MERA)~\cite{guo2024quantum}. The total circuit depth is controlled by a parameter that effectively determines the reservoir memory length, defined as the number of consecutive ansatz blocks.

A second design dimension concerns the choice of measurement-derived observables used by the classical prediction model. While prior work~\cite{yasuda2023quantum} employs single-qubit expectation values (one-point correlators), we extend the feature space to include two-point correlators in the computational basis, which can be directly estimated from measurement statistics~\cite{kornjavca2024large}. This enables us to analyze how higher-order correlations affect the expressive power of digital quantum reservoirs.

Experiments are conducted using quantum system emulators, both noiseless and noise-aware, as well as a real IQM Spark quantum processor \cite{iqm_spark}. This allows us to contrast idealized reservoir behavior with performance under realistic hardware noise and to assess the feasibility of the approach on current quantum devices.

As a concrete and operationally relevant testbed, we apply this architecture to a real-world financial time series task: forecasting daily cash demand in automated teller machines (ATMs). The dataset comprises three years of daily withdrawal amounts, and the task is formulated as rolling multi-step prediction, where each day the QRC model (composed by the QRC circuit and the linear prediction model) forecasts demand over the following 10 days. This problem exhibits non-stationary behavior and is affected by seasonality as well as stochastic fluctuations, providing a stringent benchmark for temporal modeling. Performance is compared against established classical approaches, in particular Prophet \cite{taylor2018forecasting}, a widely used state-of-the-art method for business time series forecasting.

In general, this study uses a realistic and operationally constrained forecasting problem as a testbed to evaluate a digital QRC methodology under practical design and hardware considerations. By jointly examining ansatz structure, partial measurement with reset, observable selection, and circuit depths, we assess how architectural and protocol choices translate into performance on nontrivial temporal data. This setting enables us to move beyond proof-of-principle demonstrations and provide an empirical perspective on the capabilities and limitations of digital QRC implementations on near-term quantum devices.

\section{Related works}

QRC extends the classical reservoir computing paradigm to quantum dynamical systems, exploiting high-dimensional Hilbert spaces and intrinsic quantum dynamics for temporal information processing. Similarly to classical echo state networks and liquid state machines \cite{jaeger2001echo,maass2002real, jaeger2007special}, QRC relies on a fixed nonlinear dynamical system, i.e., the reservoir, while only a linear readout layer is trained. Early foundational studies showed that NISQ devices can perform temporal processing without full fault-tolerance, establishing the feasibility of QRC for near-term hardware implementations \cite{chen2020temporal,fujii2017harnessing,nakajima2019boosting}.

Different QRC methodologies have been proposed depending on how temporal information is injected, processed, and extracted from the quantum system. One line of work focuses on measurement-driven reservoirs. Yasuda \textit{et al.} introduced a repeated-measurement framework on superconducting platforms, demonstrating that measurement back-action can enhance both memory capacity and nonlinear transformation capability \cite{yasuda2023quantum}. In this approach, measurements are interleaved with system evolution, effectively enlarging the computational space through stochastic state collapse.

A complementary direction emphasizes circuit architecture and ansatz design. Guo \textit{et al.} \cite{guo2024quantum} analyzed reusable quantum circuit patterns and abstraction principles, proposing systematic strategies to build scalable reservoirs with desirable expressivity and trainability properties. Circuit-based designs aim to engineer the dynamical properties of the reservoir through structured gate sequences.

Hardware-oriented implementations represent a third methodological axis. Kornjača \textit{et al.} demonstrated large-scale quantum reservoir learning on an analog quantum computer, achieving substantially larger reservoir sizes than typical digital implementations while maintaining stable learning performance \cite{kornjavca2024large}. Analog platforms exploit naturally occurring quantum dynamics rather than gate-based circuits, potentially offering advantages in scalability. Together, these works highlight that QRC is not a single algorithmic framework but a family of approaches differing in encoding strategies, measurement schemes, and physical realizations.

Applications of QRC to financial forecasting are relatively recent but growing. Vitali \textit{et al.} implemented QRC on a neutral-atom platform for credit card default prediction, showing performance comparable to classical machine learning baselines on a real-world classification task \cite{vitali2025quantum}. For regression-oriented time series problems, Li \textit{et al.} applied QRC to realized volatility forecasting, a task characterized by long memory and nonlinear dynamics, demonstrating that QRC models may achieve better performance than classical echo state networks under certain configurations \cite{li2025quantum}. More recently, Otieno \textit{et al.} applied QRC with few-qubit systems to nonlinear financial time series forecasting, focusing on multi-scale prediction of trading volumes in quantum-sector companies \cite{otieno2026quantum}. These studies indicate that quantum reservoirs can capture complex temporal dependencies in financial data, although most work has focused on market indicators rather than operational banking processes.

Cash demand forecasting in ATMs constitutes a distinct operational time series problem with strong practical relevance for cash logistics and inventory management. Withdrawals exhibit pronounced seasonality (daily, weekly, and yearly), calendar effects, and occasional abrupt spikes due to holidays or local events. Classical statistical approaches, for general time series prediction, include autoregressive integrated moving average (ARIMA) models and seasonal variants (SARIMA) \cite{box2015time}, as well as exponential smoothing methods such as Holt–Winters \cite{hyndman2018forecasting}. These models perform well for linear dynamics with stable seasonal structure, but may struggle with nonlinear patterns and regime changes.
Therefore, more advanced machine learning methods have been adopted for complex time series prediction. Recurrent neural networks, particularly long short-term memory (LSTM) models, can capture long-range temporal dependencies \cite{hochreiter1997long}, while tree-based ensemble methods such as gradient boosting machines provide strong performance on tabular time series features \cite{friedman2001greedy}. Hybrid approaches that combine statistical preprocessing with neural models have also been explored to improve robustness. Despite their success, these models typically require substantial training data and computational resources and may suffer from overfitting or limited interpretability.

Reservoir computing methods offer an alternative paradigm that combines nonlinear modeling capability with efficient training. Classical echo state networks have shown competitive performance in various time series tasks while maintaining low training complexity \cite{lukovsevivcius2009reservoir}. QRC inherits these advantages while potentially providing exponentially large state spaces through quantum dynamics.

Despite promising results of QRC in other financial use-cases, to the best of our knowledge, no prior work has investigated QRC for ATM cash demand forecasting. This gap motivates the present study, which explores whether quantum-enhanced reservoirs can provide advantages in modeling nonlinear temporal dependencies, long-range correlations, and robustness to noise in operational banking time series.
\section{Methodology}

Our study focuses on univariate time series forecasting using data collected from 13 ATMs. For each ATM, the dataset contains three years of daily transaction records, yielding 1,095 observations per machine. The task is to predict the cash demand for the next days based solely on past values of the same series.

For benchmarking purposes, we compare the proposed QRC approach with a classical forecasting model, \textit{Prophet} \cite{taylor2018forecasting}. Prophet is based on a generalized additive model (GAM) in which non-linear trends are decomposed into interpretable components, including seasonality (on yearly, weekly, and daily basis), as well as holiday effects. The model is designed to provide intuitive high-level parameters that can be tuned without requiring detailed knowledge of the underlying statistical formulation. To ensure a fair comparison with the proposed QRC approach, Prophet is employed in its default configuration as a preliminary benchmark, avoiding additional seasonality tuning that does not have a direct counterpart in our study.

\subsection{Preprocessing}

Both the fully classical benchmark (Prophet-based) and the QRC model share a common preprocessing stage aimed at handling inconsistencies in the raw data. Specifically, missing values (NaN or null entries) in each time series are imputed using linear interpolation based on the two nearest available observations on each side of the gap.

The cleaned time series $\{x_t\}$ is subsequently rescaled to the interval $[0,\pi]$, so that each data point can be directly encoded as input to the quantum reservoir layer. This module receives the preprocessed entries and maps them into a higher-dimensional feature space, enabling to capture complex temporal dependencies that may not be apparent in the original domain. Specifically, each rescaled value is interpreted as an angle $\theta_t$, which parameterizes the quantum operations applied to the reservoir qubits.

\subsection{Quantum reservoir layer}
The quantum reservoir layer consists of a fixed parameterized quantum circuit, built by repeating a chosen structure (ansatz), which determines the internal architecture of the reservoir. The ansatz specifies the sequence of rotation and entangling gates, thereby governing how the input-encoded angles propagate through the system and how correlations among qubits are generated. In the QRC framework, this circuit structure is chosen \textit{a priori} and remains untrained, acting as a nonlinear dynamical system that transforms the input signal into a rich set of quantum features. Different ansätze lead to different entanglement patterns, memory characteristics, and expressive capabilities, which ultimately influence the circuit’s ability to represent temporal dependencies. In this work, we consider two architectures, i.e., baseline and MERA, each providing a distinct connectivity and correlation structure for the reservoir dynamics.

\begin{figure}[htbp]
    \centering
    \scalebox{0.7}{
    \Qcircuit @C=1.0em @R=0.2em @!R { 
        \nghost{{q}_{0} :  } & \lstick{{q}_{0} :  }    & \gate{\mathrm{R_X}\,(\mathrm{\theta_t})} & \ctrl{1} & \gate{\mathrm{R_X}\,(\mathrm{\theta_t})} & \ctrl{1} & \ctrl{2} & \qw & \qw & \qw & \qw & \qw & \qw\\
        \nghost{{q}_{1} :  } & \lstick{{q}_{1} :  }    & \qw & \targ & \gate{\mathrm{R_Z}\,(\mathrm{\theta_t})} & \targ & \qw & \ctrl{2} & \qw & \qw & \qw & \qw & \qw\\
        \nghost{{q}_{2} :  } & \lstick{{q}_{2} :  }    & \qw & \qw & \qw & \qw & \targ & \qw & \meter & \gate{\mathrm{\left|0\right\rangle}} & \qw & \qw & \qw\\
        \nghost{{q}_{3} :  } & \lstick{{q}_{3} :  }    & \qw & \qw & \qw & \qw & \qw & \targ & \qw & \meter & \gate{\mathrm{\left|0\right\rangle}} & \qw & \qw\\
        \nghost{\mathrm{{c0} :  }} & \lstick{\mathrm{{c0} :  }} & \lstick{/_{_{2}}} \cw & \cw & \cw & \cw & \cw & \cw  & \dstick{_{_{\hspace{0.0em}0}}} \cw \ar @{<=} [-2,0] & \dstick{_{_{\hspace{0.0em}1}}} \cw \ar @{<=} [-1,0]& \cw  & \cw & \cw & \cw
        \gategroup{1}{6}{2}{3}{1.3em}{--} 
    }}
    \caption{Quantum circuit block for a generic $t$ time step of the baseline ansatz. System qubits $q_0$ and $q_1$ store memory across time steps and interact with auxiliary qubits $q_2$ and $q_3$, which are measured and reset at each step.}
    \label{fig:baseline_ansatz}
\end{figure}
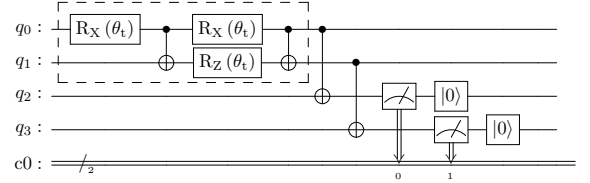

The first ansatz we consider is a \emph{baseline} model adapted from the work of \cite{yasuda2023quantum}, where it was proposed as a hardware-efficient design for their platform topology \cite{chen2020temporal}. We chose this ansatz as a starting point because our target platform, the IQM Spark, has a similar qubit connectivity, although it offers slightly greater flexibility. 
This baseline ansatz employs two system qubits that act as the memory of the reservoir over time. The system qubits undergo rotations around the $X$ and $Z$ axes, parametrized by $\theta_t$, and are entangled via CNOT gates. Following these operations, each system qubit interacts with a corresponding auxiliary qubit. The auxiliary qubits are measured and reset at every time step, allowing the reservoir to process the time series sequentially. Figure \ref{fig:baseline_ansatz} illustrates the circuit block corresponding to a single time step when using this baseline ansatz.

\begin{figure}[htbp]
    \centering
    \scalebox{0.7}{
    \Qcircuit @C=1.0em @R=0.2em @!R { 
        \nghost{{q}_{0} :  } & \lstick{{q}_{0} :  }  & \ctrl{1} & \gate{\mathrm{U}\,(\theta_t,\theta_t,\theta_t)} & \ctrl{2} & \qw & \qw & \qw & \qw & \qw & \qw\\
        \nghost{{q}_{1} :  } & \lstick{{q}_{1} :  }  & \targ & \gate{\mathrm{U}\,(\theta_t,\theta_t,\theta_t)} & \qw & \ctrl{2} & \qw & \qw & \qw & \qw & \qw\\
        \nghost{{q}_{2} :  } & \lstick{{q}_{2} :  }  & \qw & \qw & \targ & \qw & \meter & \gate{\left|0\right\rangle} & \qw & \qw & \qw\\
        \nghost{{q}_{3} :  } & \lstick{{q}_{3} :  }  & \qw & \qw & \qw & \targ & \qw & \meter & \gate{\left|0\right\rangle} & \qw & \qw\\
        \nghost{\mathrm{{c0} :  }} & \lstick{\mathrm{{c0} :  }} & \lstick{/_{_{2}}}  \cw & \cw & \cw & \cw  & \dstick{_{_{\hspace{0.0em}0}}} \cw \ar @{<=} [-2,0] & \dstick{_{_{\hspace{0.0em}1}}} \cw \ar @{<=} [-1,0]& \cw  & \cw & \cw & \cw
        \gategroup{1}{4}{2}{3}{1.1em}{--}
    }}
    \caption{MERA circuit block for a single time step $t$. The system qubits ($q_0$, $q_1$) form a hierarchical pair and interact with the auxiliary qubits ($q_2$, $q_3$). Auxiliary qubits are measured and reset after each time step, while the system qubits retain memory of previous input.}
    \label{fig:mera_ansatz}
\end{figure}
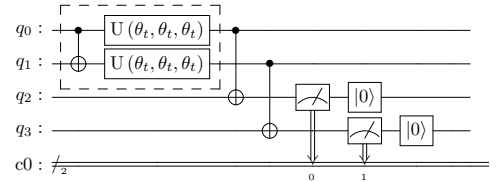

Secondly, we tested the \emph{MERA} ansatz \cite{guo2024quantum}. In MERA, each layer captures quantum entanglement at a different level, capturing dependencies with a hierarchical organization \cite{rizzi2008simulation}.
In our implementation, the two system qubits occupy different levels of the hierarchy, with $q_0$ placed at a higher level than $q_1$. A single MERA building block consists of a CNOT gate followed by a unitary operation $U(\theta_t,\theta_t,\theta_t)$, which is a single-qubit rotation defined in terms of ZYZ Euler angles \cite{qiskit_ugate}. In our scheme, one MERA block is applied to the pair of system qubits at each time step.  Then, the state of each system qubit is transferred to its corresponding auxiliary qubit via a CNOT gate. As in the baseline ansatz, the auxiliary qubits are measured and reset after each step, allowing the reservoir to process the time series sequentially. Figure~\ref{fig:mera_ansatz} illustrates the MERA-based circuit for a single time step $t$.
In this case, we also perform a final measurement on the two system qubits, providing additional observables for the reservoir readout.

\begin{figure}[htbp]
\centering
\scalebox{0.7}{
\Qcircuit @C=1.2em @R=1.0em {
\lstick{q_0:}
& \multigate{1}{\mathrm{ANSATZ}} & \ctrl{2} & \qw & \qw & \qw 
& \multigate{1}{\mathrm{ANSATZ}} & \ctrl{2} & \qw & \qw & \qw \\
\lstick{q_1:}
& \ghost{\mathrm{ANSATZ}} & \qw & \ctrl{2} & \qw & \qw 
& \ghost{\mathrm{ANSATZ}} & \qw & \ctrl{2} & \qw & \qw \\
\lstick{q_2:}
& \qw  & \targ & \qw & \meter & \gate{|0\rangle}
& \qw  & \targ & \qw & \meter & \gate{|0\rangle} \\
\lstick{q_3:}
& \qw  & \qw & \targ & \meter & \gate{|0\rangle}
& \qw  & \qw & \targ & \meter & \gate{|0\rangle} \\
\lstick{c:}
& \lstick{/_{_{2}}} \cw & \cw & \cw
& \cw \ar @{<=} [-2,0] \ar @{<=} [-1,0] & \cw
& \cw & \cw & \cw
& \cw \ar @{<=} [-2,0] \ar @{<=} [-1,0] & \cw
}
}
\caption{Two consecutive circuit blocks. Auxiliary qubits ($q_2$, $q_3$) are measured in a two-bit classical register and reset after each block.}
\label{fig:ansatz_circuit}
\end{figure}
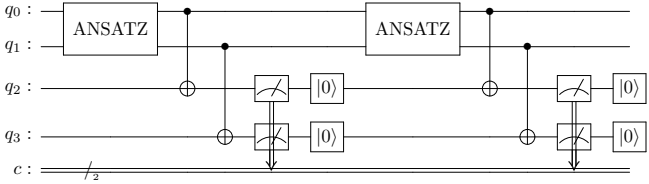

\begin{figure}[ht!]
    \centering
    \includegraphics[width=1\linewidth]{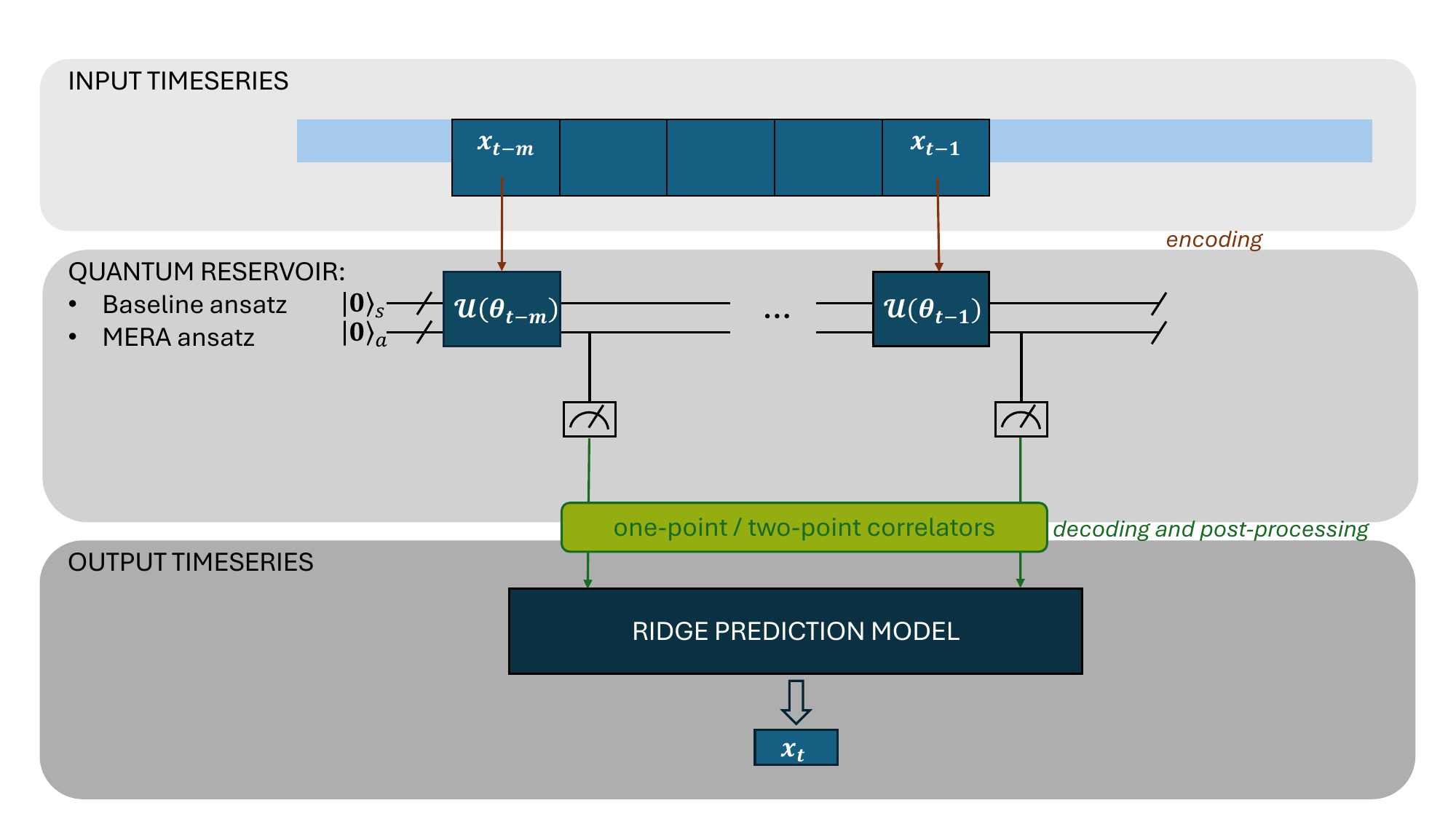}
    \caption{Illustrative scheme of the digital QRC. At each time step, $m$ consecutive time series values parameterize the quantum reservoir circuit. The resulting measurements are post-processed and fed into a forecasting model to predict $x_t$.
    }
    \label{fig:scheme}
\end{figure}

The total length of the circuit is controlled by the \textit{memory} parameter $m$, which specifies how many previous time steps are used to predict a given time step $t$. More precisely, to predict $x_t$, the scaled inputs $\theta_{t-m}, \ldots, \theta_{t-1}$ are sequentially encoded into the QRC layer through successive ansatz blocks. Figure \ref{fig:ansatz_circuit} illustrates the circuit corresponding to a single prediction with $m = 2$.
Therefore, the memory parameter determines how many consecutive days of input data are used to forecast the next day. The corresponding angles are injected into the quantum circuit, and measurements performed on the auxiliary qubits are subsequently processed by a classical linear layer to produce the final prediction, as reported in Fig. \ref{fig:scheme}.
Introducing the memory parameter provides a smooth way to control the circuit depth, allowing one to limit the impact of noise and ensure that the circuit remains within the coherence time of the hardware. This was particularly important because one of the main goals of our work was to test the approach on the IQM Spark QPU. Furthermore, the parameter enables testing memory values that are consistent with the temporal granularity of the application. In our case, we considered values of $m$ that are multiples of 7 to capture the most basic weekly periodicity in the daily time series data.

\subsection{Regression}\label{sec:regression}

Given the measurement outcomes on the auxiliary qubits (and on the system qubits at the last time step, in the case of the MERA circuit), one can extract different observables from the resulting bitstrings. In particular, we consider one- and two-point correlators, which act as features for the downstream linear prediction model.

Let $n = 2m$ ($n = 2m + 2$ in the case of the MERA circuit) be the number of measurements per shot, and let $\mathcal{D} = \{(s_j, c_j)\}$ denote the measurement dataset, where $s_j \in \{0,1\}^n$ is a bitstring outcome and $c_j$ is the number of times it was observed, with $j$ indexing the distinct observed bitstrings. The total number of shots is $N = \sum_{j} c_j$.

\paragraph{Single-qubit observables}
The expectation value of the Pauli-$Z$ operator on qubit $i$ is estimated from the measurement statistics as
\begin{equation}
\langle Z_i \rangle
= \sum_{j} \frac{c_j}{N} \, (-1)^{s_j^{(i)}},
\end{equation}
where $s_j^{(i)} \in \{0,1\}$ is the $i$-th bit of the string $s_j$. 

Collecting all single-qubit expectations yields the feature vector
\begin{equation}
\mathbf{z}^{(1)} =
\big( \langle Z_0 \rangle, \ldots, \langle Z_{n-1} \rangle \big)
\in \mathbb{R}^n .
\end{equation}

\paragraph{Two-qubit correlators}
To capture pairwise correlations, we also compute expectation values of products of Pauli-$Z$ operators. For a pair of qubits $(i,k)$,
\begin{equation}
\langle Z_i Z_{k} \rangle
= \sum_{j} \frac{c_j}{N} \, (-1)^{\, s_j^{(i)} + s_j^{(k)}} .
\end{equation}
In our implementation, we evaluate this quantity for all possible pairs of qubits. This yields the feature vector

\begin{equation}
\mathbf{z}^{(2)} =
\big(
\langle Z_0 Z_1 \rangle,\,
\langle Z_0 Z_2 \rangle,\,
\ldots,\,
\langle Z_{n-2} Z_{n-1} \rangle
\big)
\in \mathbb{R}^{n(n-1)/2}.
\end{equation}

In our experiments, we considered two feature configurations:

\begin{itemize}
    \item \textit{Single-qubit features only:} the regression model receives $\mathbf{f} =\mathbf{z}^{(1)}$ as input.
    \item \textit{Augmented features:} the input consists of the concatenation of single-qubit observables and two-qubit correlators, $\mathbf{f} = \big(\mathbf{z}^{(1)}, \mathbf{z}^{(2)} \big).$
\end{itemize}

The second configuration was introduced because preliminary experiments indicated that single-qubit observables alone were insufficient to capture the complex temporal dependencies present in the time series. Including two-point correlators enriches the feature space with information about quantum correlations between qubits, thereby enabling the linear model to represent more expressive nonlinear transformations of the input data.

At each time step $t$, the feature vector $\mathbf{f}_t$ is computed from the measurements of the quantum circuits, encoding the input $x_{t-m}, \ldots, x_{t-1}$.
The predicted value at time step $t$ is obtained using a regularized linear regression model. In our experiments, we used the \textit{Ridge} estimator from the \texttt{scikit-learn} library \cite{pedregosa2011scikit, hao2019machine}, which applies a penalty $\ell_2$ to the regression coefficients. Given a sequence of feature vectors $\{\mathbf{f}_t\}_{t=1}^{T}$ extracted from the quantum circuit and corresponding target values $\{x_t\}_{t=1}^{T}$, Ridge regression solves

\begin{equation}\label{eq:ridge}
\min_{\mathbf{w},\,b} 
\frac{1}{2T} \sum_{t=1}^{T} \left( x_t - \mathbf{w}^\top \mathbf{f}_t - b \right)^2
\;+\; \frac{\alpha}{2}\,\|\mathbf{w}\|_2^2,
\end{equation}

where $\alpha=1.0$ is the regularization strength controlling the amount of shrinkage applied to the coefficients. The fitted parameters $(\mathbf{w},b)$ then define the prediction at time step $t$ as

\begin{equation}
\hat{x}_t = \mathbf{w}^\top \mathbf{f}_t + b.
\end{equation}

To ensure well-defined inputs, we discard the first $m$ time steps from the training set, since the construction of each feature vector $\mathbf{f}_t$ depends on $m$ preceding observations, which are unavailable for these initial time steps.

Finally, the overall QRC model operates in a closed-loop configuration to enable multi-day-ahead predictions. Although the regression model is trained to predict one day ahead, we recursively feed predictions as input to the QRC to generate new features for the classical model, thereby enabling forecasts up to a 10-day horizon. 
Since this recursive strategy can accumulate errors, we also experimented with a multi-output regressor implemented as a \textit{RegressorChain} with Ridge base estimators. In this approach, the $h$-th output is predicted using a Ridge model that takes as input both the original feature vector $\mathbf{f}_t$ and the predictions of the previous $h-1$ outputs,
\begin{equation}
\hat{x}_t^{(h)} =
\mathbf{w}_h^\top \tilde{\mathbf{f}}_t^{(h)} + b_h,
\end{equation}
where
\begin{equation}
\tilde{\mathbf{f}}_t^{(h)} =
\begin{bmatrix}
\mathbf{f}_t \\
\hat{x}_t^{(1)} \\
\vdots \\
\hat{x}_t^{(h-1)}
\end{bmatrix}.
\end{equation}

Note that $\tilde{\mathbf{f}}_t^{(1)} = \mathbf{f}_t$. Each model in the chain is trained independently by minimizing the Ridge objective
\begin{equation}
\min_{\mathbf{w}_h, b_h}
\frac{1}{2T}\sum_{t=1}^{T}
\left(x_t^{(h)} -
\mathbf{w}_h^\top \tilde{\mathbf{f}}_t^{(h)}
- b_h \right)^2
+ \frac{\alpha}{2}\|\mathbf{w}_h\|_2^2 .
\end{equation}

\section{Results}

The QRC circuits are constructed and simulated using the \texttt{quantumreservoirpy} library \cite{miao2024quantumreservoirpy}, which supports the closed-loop implementation required for our task. It is a Qiskit-based Python library that allows one to define the reservoir ansatz and, if needed, override internal methods to compute custom observables as input to the regression model. This flexibility enabled seamless testing across multiple configurations. Specifically, we explored the following settings:
\begin{itemize}
    \item \textbf{Ansatz:} baseline and MERA,
    \item \textbf{Memory} $m$: 7, 14, 21, 28, 35, 42,
    \item \textbf{Input features} $\mathbf{f}$: single- and two-point correlators,
    \item \textbf{Backend:} Aer Simulator \cite{qiskit_aer} (noiseless), IQM Adonis \cite{iqm_adonis, iqm_noise_modeling} (noise model), and IQM Spark \cite{iqm_spark} (real QPU).
\end{itemize}

The feature vectors $\mathbf{f}_t$ are calculated using $N = 100$ measurement shots at each time step $t$. The prediction quality is then evaluated using three complementary metrics:

\begin{itemize}
    \item Normalized Mean Squared Error (NMSE):
    \begin{equation}
    \mathrm{NMSE}
    =
    \frac{\sum_{t} (x_t - \hat{x}_t)^2}
         {\sum_{t} (x_t - \bar{x})^2},
    \end{equation}
    where $\bar{x}$ is the mean of the true values. NMSE measures the relative squared error with respect to the variance of the data, allowing comparison between time series with different scales.

    \item Mean Absolute Error (MAE):
    \begin{equation}
    \mathrm{MAE}
    =
    \frac{1}{T}
    \sum_{t=1}^{T}
    |x_t - \hat{x}_t|,
    \end{equation}
    which provides an interpretable average magnitude of the prediction error and is less sensitive to outliers than MSE.

    \item Dynamic Time Warping (DTW): 
    DTW computes the minimal alignment cost between predicted and true sequences under nonlinear temporal warping \cite{senin2008dynamic, JMLR:v21:20-091}. Given two sequences $\mathbf{x}=(x_1,\ldots,x_T)$ and $\hat{\mathbf{x}}=(\hat{x}_1,\ldots,\hat{x}_T)$, DTW is defined as
    \begin{equation}
    \mathrm{DTW}(\mathbf{x}, \hat{\mathbf{x}})
    =
    \min_{p}
    \sum_{(i,j)\in p}
    d(x_i,\hat{x}_j),
    \end{equation}
    where $p$ is a warping path satisfying boundary, monotonicity, and continuity constraints, and $d(\cdot,\cdot)$ is a local distance measure $d(a,b)=(a-b)^2$. Unlike pointwise metrics such as NMSE or MAE, DTW accounts for small temporal shifts between sequences, and therefore captures similarity in overall shape rather than strict point-to-point correspondence.
\end{itemize}

Since each prediction is performed over a time window of $K=10$ days, at every time step $t$ the entire model (including the QRC and the regression modules) returns forecasts for the following $K$ days. Consequently, performance metrics can be averaged along two possible dimensions: either across the $K$ time series, each having a test set of length $T$, or across the $T$ time steps, each encompassing $K$ time series.

\subsection{Results on ATM 1 time series}

\begin{figure*}[ht!]
    \centering
    \begin{subfigure}{0.88\linewidth}
        \centering
        \includegraphics[width=\linewidth]{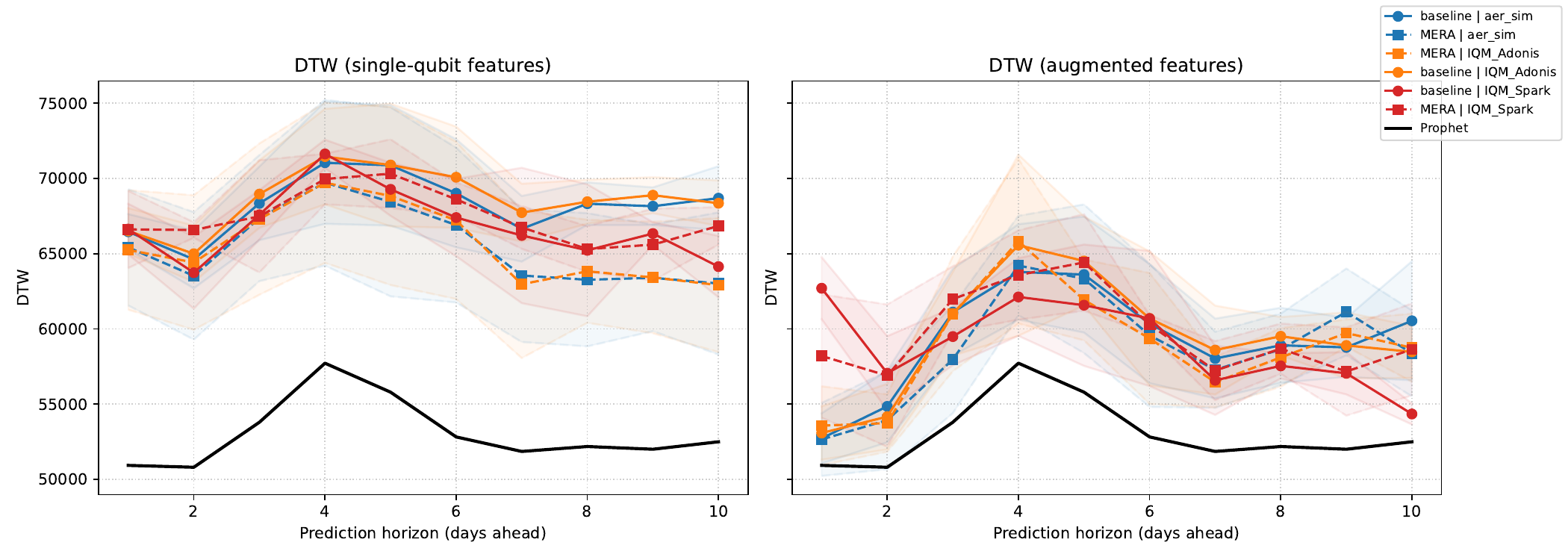}
        \caption{}
        \label{fig:dtw_h}
    \end{subfigure}
    \hfill
    \begin{subfigure}{0.88\linewidth}
        \centering
        \includegraphics[width=\linewidth]{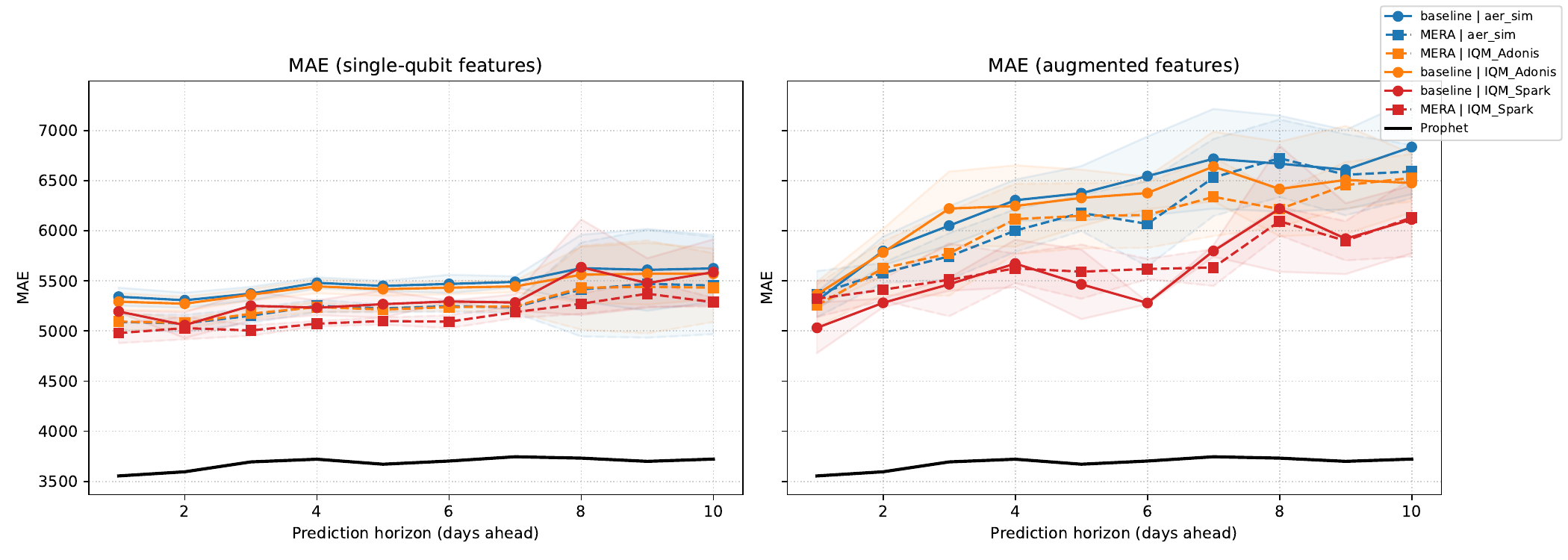}
        \caption{}
        \label{fig:mae_h}
    \end{subfigure}
    \hfill
    \begin{subfigure}{0.88\linewidth}
        \centering
        \includegraphics[width=\linewidth]{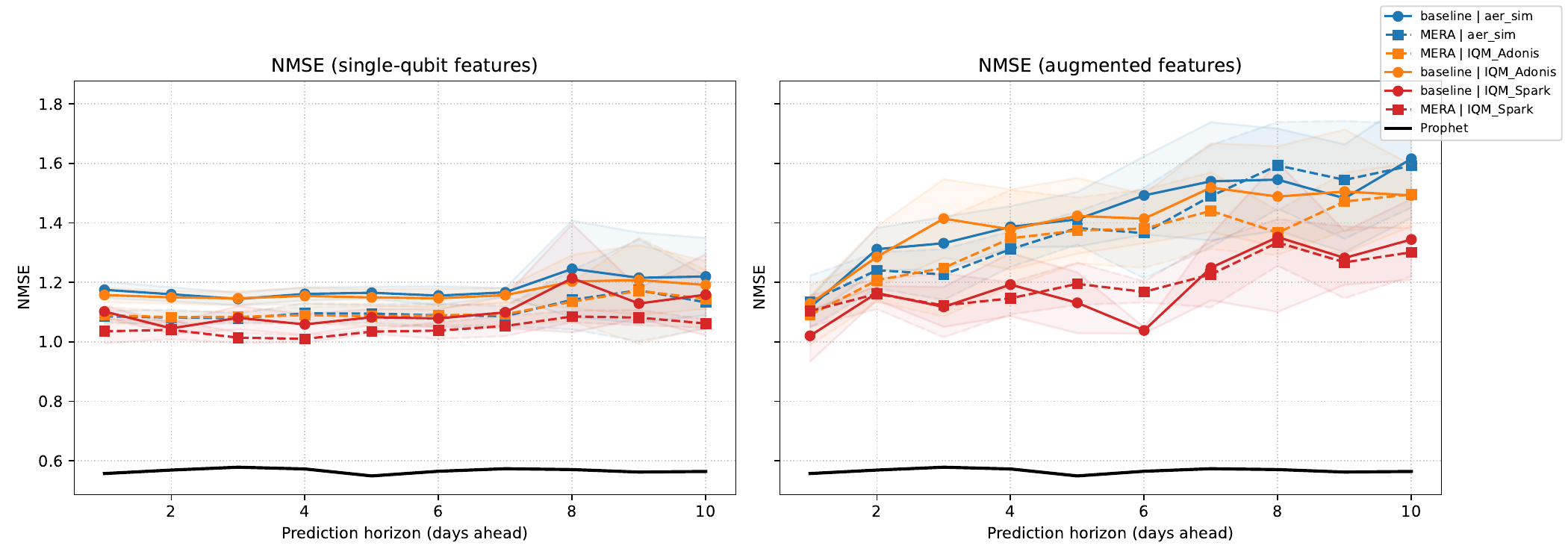}
        \caption{}
        \label{fig:nmse_h}
    \end{subfigure}
    
    \caption{Performance metrics as a function of the prediction horizon ($h=1,\ldots,10$). The semi-transparent region around each curve represents one standard deviation from the mean for each value of $h$.}
\end{figure*}

\begin{figure*}[ht!]
    \centering
    \begin{subfigure}{0.88\linewidth}
        \centering
        \includegraphics[width=\linewidth]{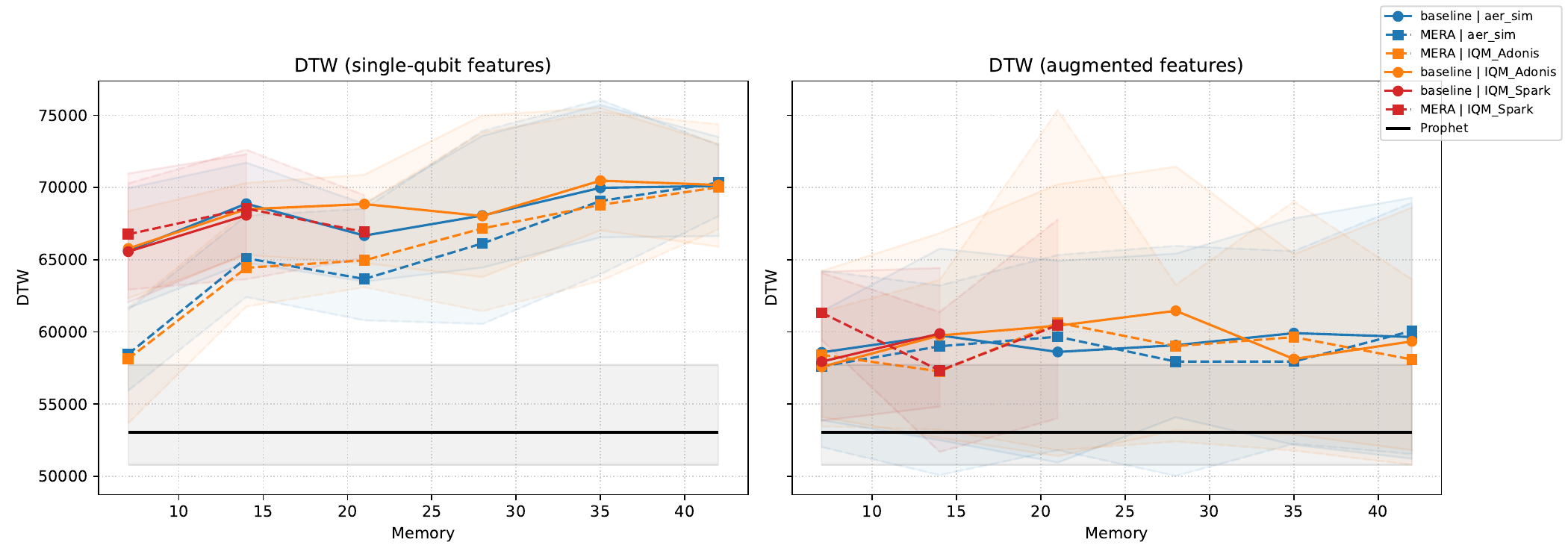}
        \caption{}
        \label{fig:dtw_m}
    \end{subfigure}
    \hfill
    \begin{subfigure}{0.88\linewidth}
        \centering
        \includegraphics[width=\linewidth]{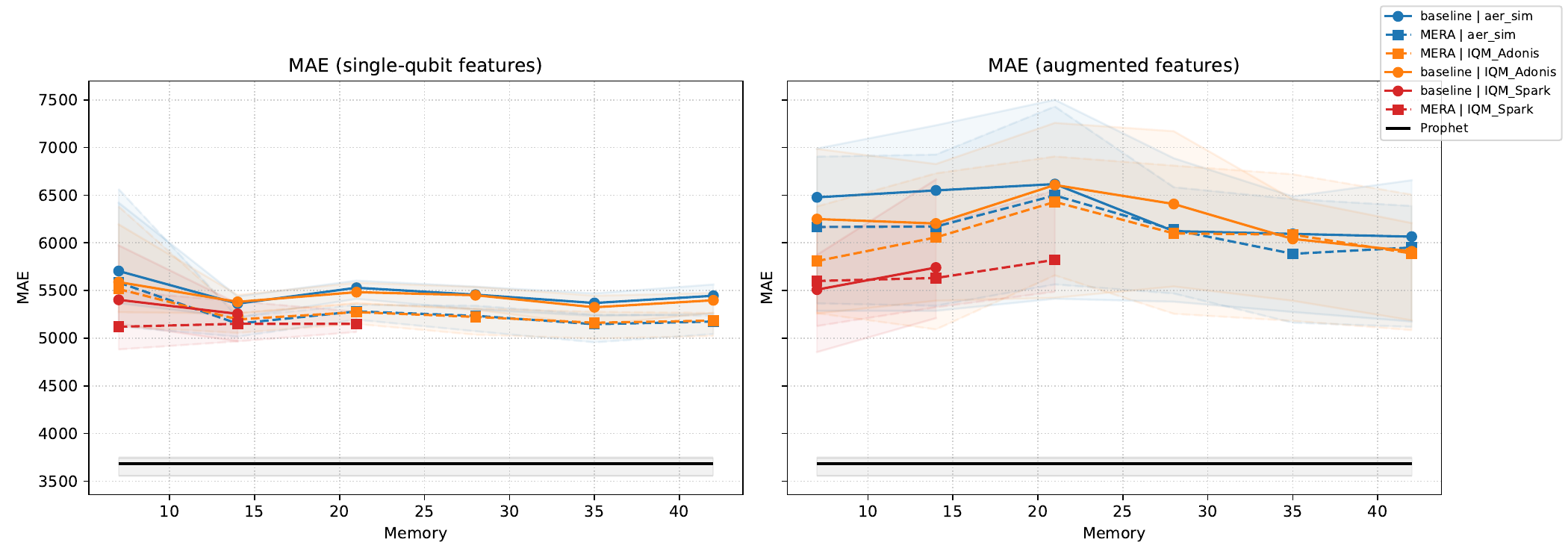}
        \caption{}
        \label{fig:mae_m}
    \end{subfigure}
    \hfill
    \begin{subfigure}{0.88\linewidth}
        \centering
        \includegraphics[width=\linewidth]{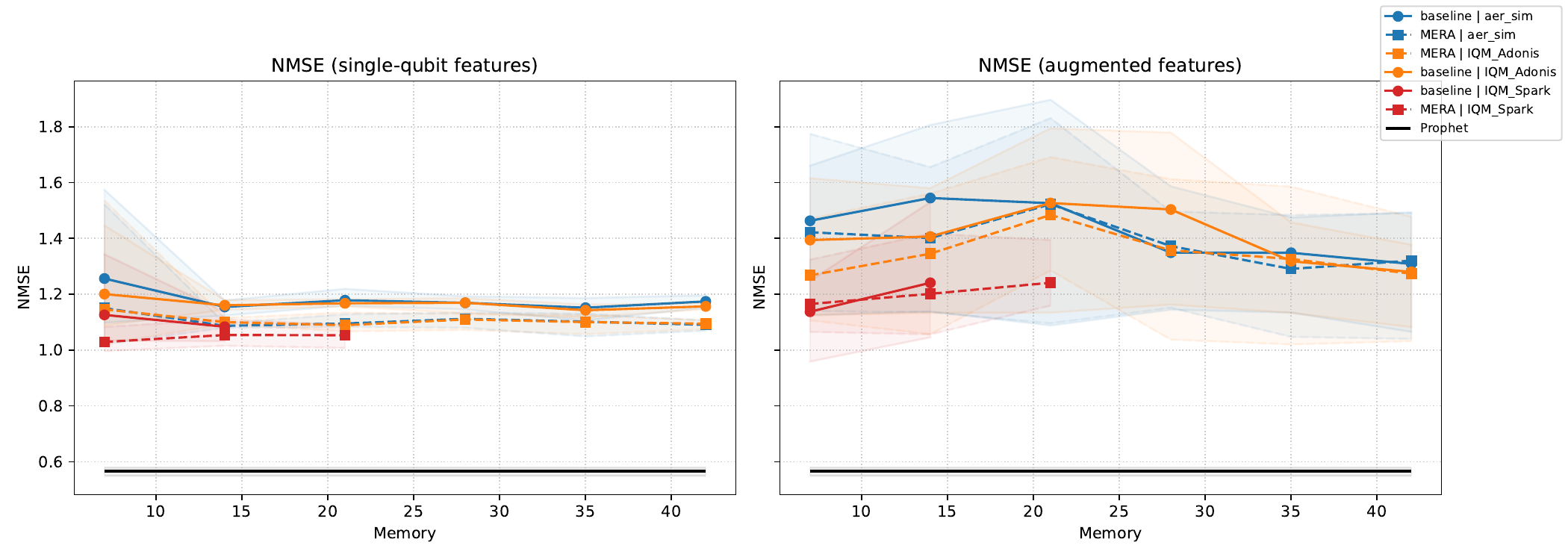}
        \caption{}
        \label{fig:nmse_m}
    \end{subfigure}
    
    \caption{Performance metrics as a function of the QRC memory parameter. The metrics are averaged across the prediction horizons. The shaded regions represent the minimum and maximum values obtained in the tested memory configurations. Due to hardware execution time constraints, the memory values $m \geq 28$ (MERA ansatz) and $m \geq 21$ (baseline ansatz) are not included.}
\end{figure*}

\begin{figure*}[ht!]
    \centering
    \begin{subfigure}{0.88\linewidth}
        \centering
        \includegraphics[width=\linewidth]{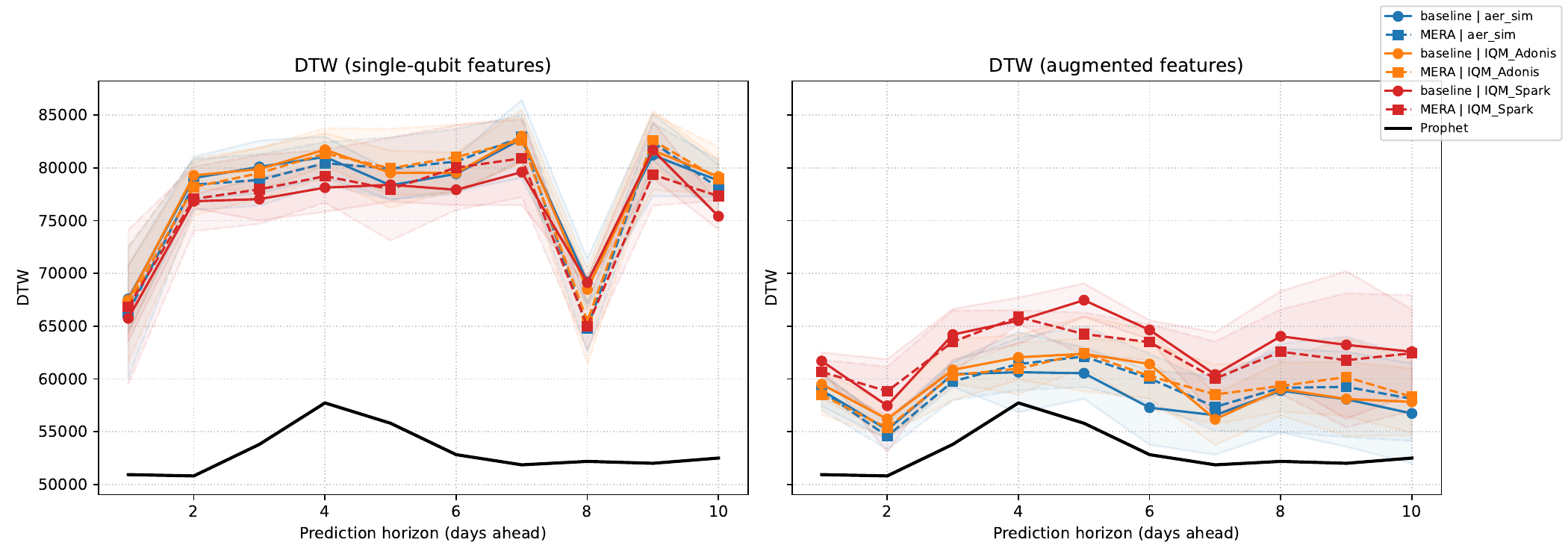}
        \caption{}
        \label{fig:dtw_h_10}
    \end{subfigure}
    \hfill
    \begin{subfigure}{0.88\linewidth}
        \centering
        \includegraphics[width=\linewidth]{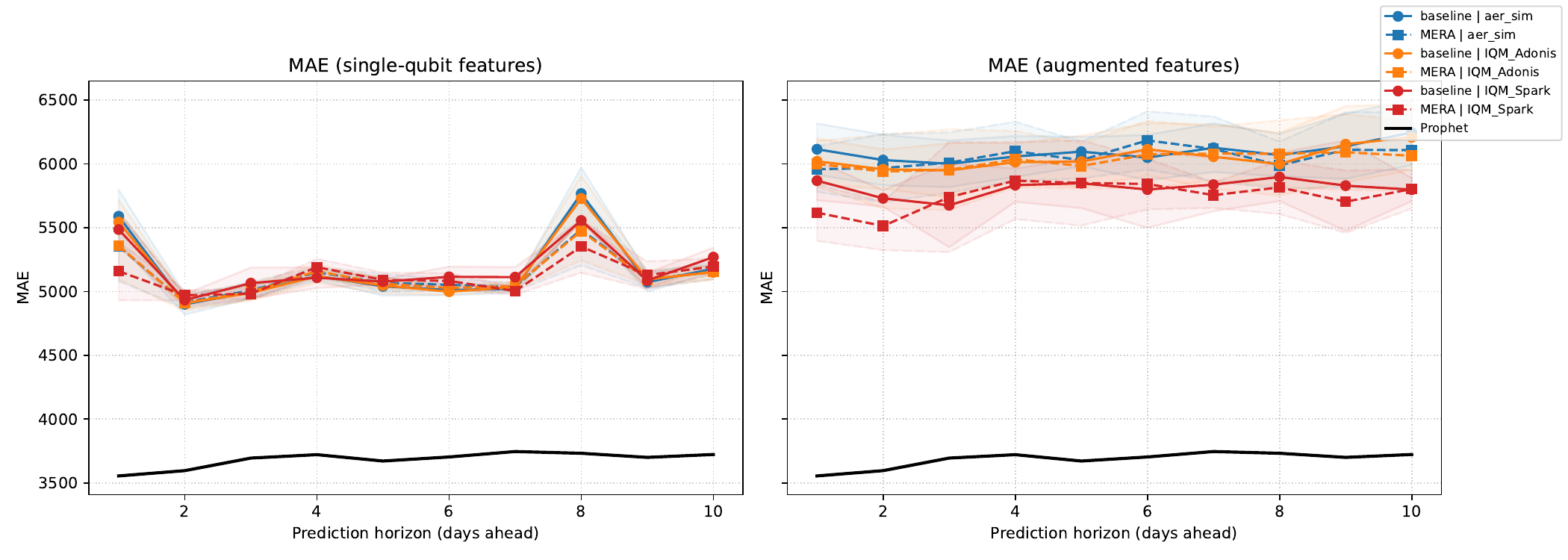}
        \caption{}
        \label{fig:mae_h_10}
    \end{subfigure}
    \hfill
    \begin{subfigure}{0.88\linewidth}
        \centering
        \includegraphics[width=\linewidth]{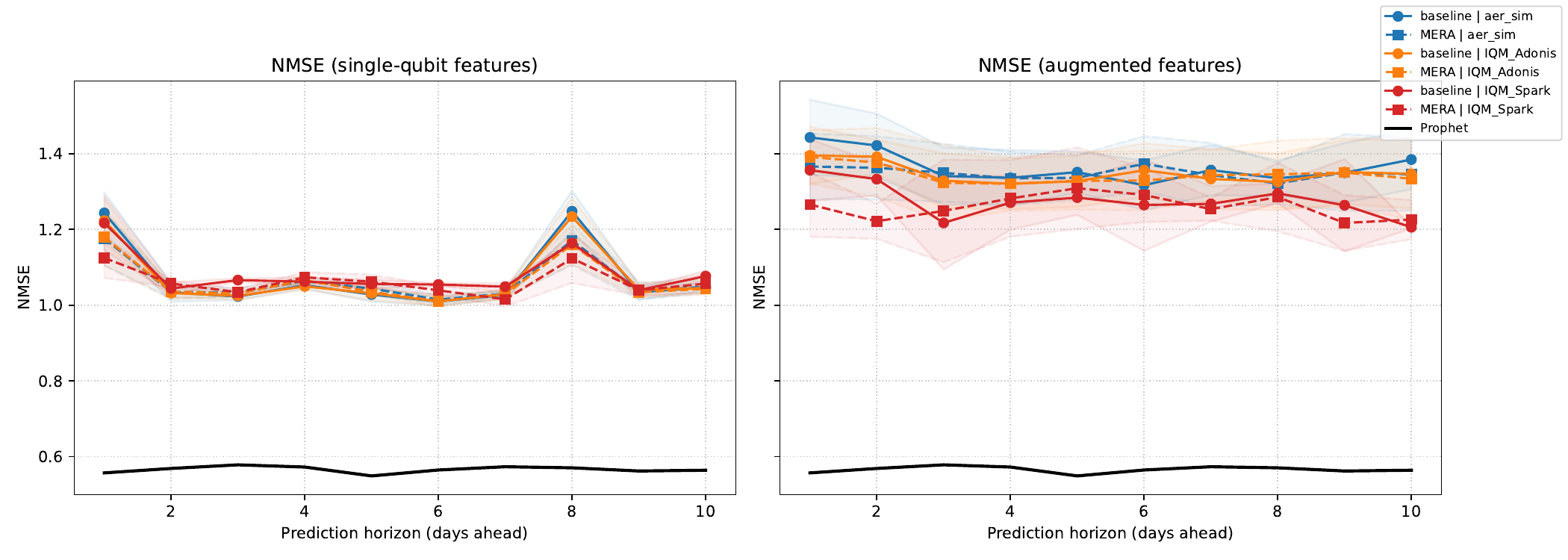}
        \caption{}
        \label{fig:nmse_h_10}
    \end{subfigure}
    
    \caption{Performance metrics as a function of the prediction horizon ($h=1,\ldots,10$), obtained using a Ridge regressor chain model after the QRC module. The semi-transparent region around each curve represents one standard deviation from the mean.}
\end{figure*}

We present the complete study on all configurations of the framework for the first ATM (ATM 1). The time series, which spans three years, is split such that the first two and a half years are used to train the Ridge model weights \eqref{eq:ridge}, while the final half year ($T = 172$) serves as the test set.

In each time step, the preceding $m$ data points are used to compute the feature vector $\mathbf{f}_t$, and the framework predicts a 10-day window (the next ten data points). The Ridge model is retrained at each prediction step using a sliding-window strategy: after each forecast, the training set is advanced by one day, and the model is refitted until only the final 10 days remain in the test set. 
Because each prediction produces a 10-day horizon ($K=10$), the test set has an additional dimension corresponding to the forecast lead time. This results in ten predicted time series $\{\hat{x}_t^{(h)}\}_{h=1}^{K}$, each associated with a specific number of days ahead $h$. 

The following results are obtained using a Ridge regressor (Eq. \eqref{eq:ridge}) operating in a closed-loop configuration with the QRC module.

The first analysis is obtained by evaluating the forecasting metrics separately for each prediction horizon. 
Specifically, the predicted value $\hat{x}_t^{(h)}$ is compared with the corresponding ground-truth value $x_{t+h}$. 
Figures~\ref{fig:dtw_h}, \ref{fig:mae_h}, and \ref{fig:nmse_h} show the DTW, MAE, and NMSE metrics, respectively, 
computed for each time series at the horizons $h = 1,2,\ldots,10$.
The results are reported for different backends used for the quantum reservoir execution: noiseless simulation 
(\texttt{aer\_sim}), noisy emulation (\texttt{IQM\_Adonis}), and executions on quantum hardware 
(\texttt{IQM\_Spark}). We evaluated both the baseline ansatz and the MERA architecture.
For the QRC module, multiple memory configurations are tested. The plots report the average metric across the different memory values ($m=$ 7, 14, 21, 28, 35, 42). As a classical reference, we also include the benchmark obtained using Prophet.

From these plots, we observe that, for both the MAE (Fig. \ref{fig:mae_h}) and NMSE (Fig. \ref{fig:nmse_h}) metrics, the Prophet model exhibits a largely constant performance across the prediction horizons. This behavior is expected, as Prophet is specifically designed for multi-step forecasting over longer temporal ranges.
In contrast, QRC-based methods generally perform worse than the Prophet benchmark and show an increasing error as the prediction horizon grows, with the largest errors occurring at longer horizons. This trend is particularly evident when augmented features are used to train the Ridge regression readout.
A possible explanation for this behavior is the accumulation of forecast errors in subsequent prediction steps, which progressively degrades the accuracy of the predicted values on longer horizons.

For the DTW metric (Fig. \ref{fig:dtw_h}), we observe a different behavior. In this case, neither the Prophet model nor the QRC-based models exhibit a constant trend across the prediction horizons. Instead, the DTW score tends to increase during the first four days of the horizon and then decreases for longer horizons.
For this metric, the performance gap with respect to the Prophet benchmark is smaller, especially when using augmented features. In some cases, the QRC methods even achieve slightly better performance than Prophet during the first two days of the prediction horizon.

The second analysis is performed by examining metrics across different backends, ansatz architectures, QRC feature types, and memory values. In this case, the metrics are averaged over the prediction horizons.
This analysis aims to determine whether a particular configuration provides consistently better performance. We observe that the augmented-feature models exhibit higher variability compared to the single-qubit feature models, especially for larger memory values.
Similarly to horizon-based analysis, the Prophet benchmark remains unbeaten for the MAE (Fig.~\ref{fig:mae_m}) and NMSE (Fig.~\ref{fig:nmse_m}) metrics. However, for the DTW metric (Fig.~\ref{fig:dtw_m}), there are configurations in which the QRC models achieve slightly better performance than the Prophet baseline.

From these results, particularly when considering the MAE (Fig. \ref{fig:mae_m}) and NMSE (Fig. \ref{fig:nmse_m}) metrics, it appears that the noise introduced by both the simulated noise model (\texttt{IQM\_Adonis}) and, more significantly, the real hardware (\texttt{IQM\_Spark}), actually improves the quality of the predictions. One possible interpretation is that noise injects additional nonlinearities into the quantum reservoir, thereby enhancing its ability to capture complex relationships between temporal features.
However, an important limitation of using the real QPU should be noted: circuits with execution times longer than $800$ ms cannot be run. In our case, this constraint prevents us from testing memory values of $m \geq 28$ for the MERA ansatz and $m \geq 21$ for the baseline ansatz.

While assessing whether a multi-output model could be less affected by the accumulation of quantum noise over increasing prediction horizons, we tested the Regressor Chain model described in Section~\ref{sec:regression}, repeating all combinations explored in the previous analysis.
As shown in Figures~\ref{fig:dtw_h_10}--\ref{fig:nmse_h_10}, the chain model does not improve the convergence of the considered metrics. Moreover, inspecting the predicted time series for some of the most promising configurations, we observe that, for certain prediction horizons, the model produces outputs that exhibit significantly reduced variability compared to the true series. This overly smooth behavior suggests a loss of temporal dynamics. Consistently, the narrower standard deviation bands indicate reduced variability across runs, likely reflecting a tendency of the model to converge toward similar, smoothed predictions.
For this reason, we revert to the single-regressor model which, although still underperforming the Prophet benchmark, does not exhibit the overly smooth behavior observed with the chain model. 

\begin{figure}
    \centering
    \begin{subfigure}{\linewidth}
        \centering
        \includegraphics[width=\linewidth]{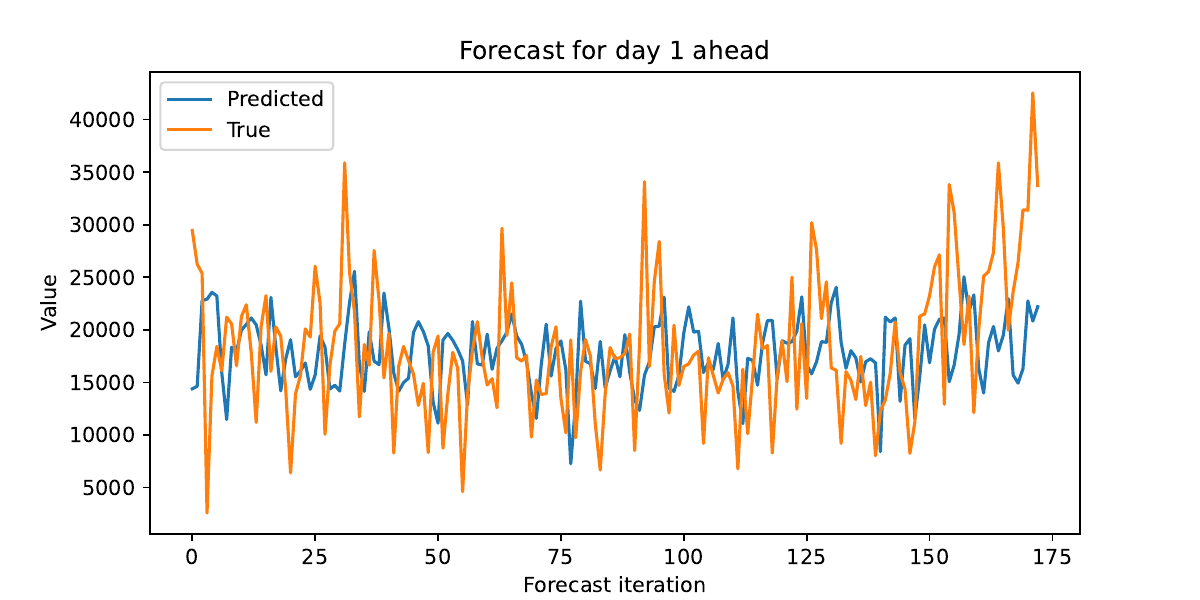}
        \caption{QRC (MERA, $m=21$)}
    \end{subfigure}
    \hfill
    \begin{subfigure}{\linewidth}
        \centering
        \includegraphics[width=\linewidth]{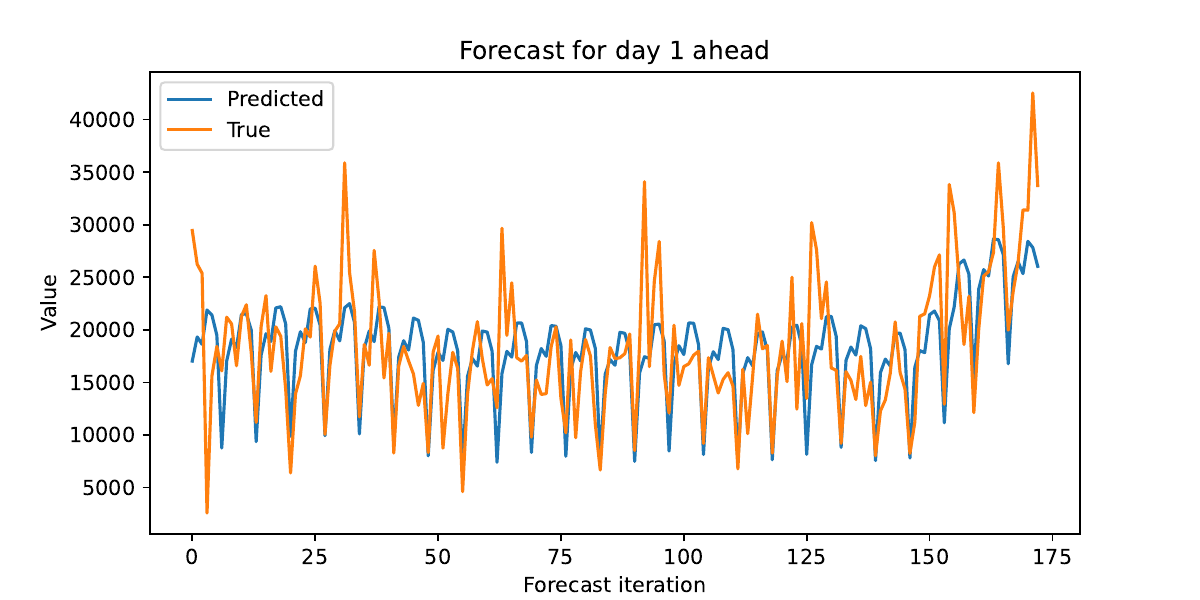}
        \caption{Prophet}
    \end{subfigure}
    
    \caption{Comparison of one-step-ahead ($h=1$) predictions for the best QRC configuration and the Prophet benchmark.}
    \label{fig:ts_comparison_h1}
\end{figure}

The best-performing configuration compatible with the IQM Spark QPU employs augmented features, the MERA ansatz, and a memory parameter $m = 21$, with detailed averaged metric values reported in Tables \ref{tab:mae_atms}–\ref{tab:dtw_atms}. Figure~\ref{fig:ts_comparison_h1} compares the time series predicted by Prophet and the QRC framework for a one-step-ahead forecast ($h=1$).

\subsection{Best QRC configurations applied to the other ATMs}

In this section, based on the analysis performed on the ATM 1 time series, we select the most promising QRC configuration (augmented features, MERA ansatz, and $m = 21$ as memory parameter) and assess its generalization capability by applying it to the remaining ATMs. 
The same training and evaluation protocol described in the previous section is adopted. In particular, for each ATM, the model is trained using a sliding-window approach and evaluated over a prediction horizon of $K=10$ days. Forecast performance is measured using the DTW, MAE, and NMSE metrics compared with the Prophet benchmark.

Tables~\ref{tab:mae_atms}--\ref{tab:dtw_atms} report the forecast performance of the selected QRC configuration across the remaining ATMs. In general, the results are consistent with those observed for ATM~1.

From Table~\ref{tab:mae_atms}, the MAE values obtained with QRC are systematically higher than those achieved by Prophet across all ATMs, although a consistent improvement can be observed when moving from the simulator to hardware-inspired noise models, and further to the IQM Spark backend. A similar trend is observed for the NMSE values in Table~\ref{tab:nmse_atms}, where the QRC exhibits significantly larger errors, indicating difficulties in capturing the variance of the underlying time series.

The DTW metric reported in Table~\ref{tab:dtw_atms} provides a more nuanced picture. Although QRC generally underperforms Prophet, for certain ATMs (e.g., ATM~3 and ATM~13) the QRC model with the IQM Spark backend achieves better performance than Prophet, indicating its ability to capture temporal patterns despite inaccuracies in amplitude.

Overall, these results indicate that, despite the selected QRC configuration yields a consistent behavior across different ATMs, it does not outperform the classical benchmark. The observed improvements when incorporating more realistic hardware settings suggest that noise and hardware-specific effects may play a non-trivial role in shaping the QRC model performance.

\begin{table}[ht]
\centering
\caption{Performance of the selected QRC configuration (augmented features, MERA, $m=21$) across ATMs, compared with Prophet.}
\label{tab:performance_atms}

\begin{subtable}{\columnwidth}
\centering
\caption{MAE}
\begin{tabular}{c|ccc|c}
\hline
\textbf{ATM ID} & aer\_sim & IQM\_Adonis & IQM\_Spark & Prophet \\
\hline
1  & 6496.96 & 6429.04 & 5821.56 & 3684.08 \\
2  & 6274.29 & 6237.80 & 5884.80 & 2572.45 \\
3  & 3108.59 & 2945.22 & 2859.91 & 1695.34 \\
4  & 3674.64 & 3522.65 & 3301.53 & 2145.67 \\
5  & 15091.12 & 14050.05 & 11871.03 & 5495.49 \\
6  & 8284.42 & 8765.06 & 6191.27 & 3466.30 \\
7  & 12857.05 & 12093.33 & 11525.46 & 7018.47 \\
8  & 2051.72 & 1946.58 & 1821.04 & 1057.47 \\
9  & 8883.56 & 8489.53 & 7897.17 & 4718.06 \\
10 & 4358.33 & 4375.25 & 4273.41 & 2823.99 \\
11 & 1613.32 & 1555.69 & 1440.70 & 1169.06 \\
12 & 1243.94 & 1023.17 & 878.48  & 615.58 \\
13 & 6672.61 & 6534.34 & 4156.20 & 3088.79 \\
\hline
\end{tabular}\label{tab:mae_atms}
\end{subtable}

\vspace{0.5cm}

\begin{subtable}{\columnwidth}
\centering
\caption{NMSE}
\begin{tabular}{c|ccc|c}
\hline
\textbf{ATM ID} & aer\_sim & IQM\_Adonis & IQM\_Spark & Prophet \\
\hline
1  & 1.52 & 1.48 & 1.24 & 0.57 \\
2  & 1.36 & 1.33 & 1.14 & 0.24 \\
3  & 1.44 & 1.30 & 1.20 & 0.46 \\
4  & 1.63 & 1.48 & 1.32 & 0.68 \\
5  & 1.49 & 1.30 & 0.95 & 0.31 \\
6  & 3.19 & 3.67 & 1.80 & 0.74 \\
7  & 1.77 & 1.55 & 1.37 & 0.70 \\
8  & 1.72 & 1.58 & 1.29 & 0.52 \\
9  & 1.08 & 1.00 & 0.83 & 0.37 \\
10 & 1.31 & 1.31 & 1.26 & 0.64 \\
11 & 1.53 & 1.44 & 1.23 & 0.85 \\
12 & 3.71 & 2.47 & 1.70 & 0.95 \\
13 & 3.38 & 3.22 & 1.34 & 1.10 \\
\hline
\end{tabular}\label{tab:nmse_atms}
\end{subtable}

\vspace{0.5cm}

\begin{subtable}{\columnwidth}
\centering
\caption{DTW}
\begin{tabular}{c|ccc|c}
\hline
\textbf{ATM ID} & aer\_sim & IQM\_Adonis & IQM\_Spark & Prophet \\
\hline
1  & 59666.21 & 60634.84 & 60468.08 & 53036.60 \\
2  & 45360.68 & 45209.23 & 45154.36 & 34525.37 \\
3  & 25791.76 & 24739.83 & \textbf{24492.86} & \textbf{25069.78}\\
4  & 35051.77 & 34481.09 & 34543.40 & 33755.24 \\
5  & 115860.03 & 109111.60 & 95364.39 & 92187.12 \\
6  & 72500.54 & 80262.73 & 56233.78 & 54591.18 \\
7  & 110053.39 & 106243.34 & 110047.63 & 105359.84 \\
8  & 17039.28 & 16194.22 & 17157.77 & 15032.21 \\
9  & 77148.10 & 75237.93 & 73336.58 & 72808.60 \\
10 & 38908.17 & 38729.23 & 39481.31 & 37088.30 \\
11 & 13611.34 & 13288.43 & 14125.26 & 13979.97 \\
12 & 10283.80 & 8743.95  & 7977.51  & 7703.04 \\
13 & 55769.59 & 54478.13 & \textbf{40517.00} & \textbf{43294.14} \\
\hline
\end{tabular}\label{tab:dtw_atms}
\end{subtable}

\end{table}
\section{Conclusion}

In this work, we investigated a digital QRC framework for multi-step forecasting of ATM cash demand time series. The study focused on how architectural and protocol choices in the quantum reservoir, namely the ansatz structure, the choice of measurement-derived observables, the memory parameter, and the execution backend, affect forecasting performance in a realistic financial prediction task.

Our results show that the proposed QRC approach is capable of producing meaningful forecasts and capturing part of the temporal structure of the data, especially when evaluated through the DTW metric, which is less sensitive to pointwise amplitude mismatches. In particular, configurations based on augmented features, combining one- and two-point correlators, generally yielded the best performance, indicating that higher-order measurement information improves the expressive power of the reservoir readout. Among the tested architectures, the MERA ansatz provided the most promising behavior, and the configuration that uses augmented features with MERA and memory $m=21$ emerged as the best compromise between predictive quality and hardware feasibility on the IQM Spark platform.
At the same time, the QRC models did not outperform the Prophet benchmark on MAE and NMSE. This indicates that, for the considered dataset and training protocol, the quantum reservoir is still less effective than a strong classical forecasting baseline in reproducing accurate pointwise predictions and variance levels. Nevertheless, the smaller gap observed in DTW, and occasional outperformance, suggests that the QRC can partially capture the temporal evolution and overall shape of the series, even when the prediction amplitudes remain less accurate.

An interesting outcome of the study is the role played by realistic hardware effects. Moving from noiseless simulation to noise-aware emulation, and in several cases further to real-hardware execution, often led to improved performance. Although this effect should be interpreted cautiously, it suggests that noise and hardware-specific dynamics may contribute nontrivially to the reservoir transformation, potentially improving the effective feature map available to the classical readout \cite{domingo2023taking}. At the same time, hardware execution limits the maximum accessible circuit depth and therefore the range of memory values that can be explored. Moreover, since noise is hardware-specific, the performance gains may not generalize across different devices, instead require tailoring the QRC model to a given platform. These aspects highlight a trade-off between leveraging quantum noise and maintaining sufficient circuit expressivity and robustness across hardware implementations. Future work will aim to better understand this interplay by testing different QPUs and exploring more complex ansatz\"e, with the goal of identifying optimal QRC configurations in a wider range of settings.

\section*{Acknowledgment}
The authors thank the domain Gestione Integrata Dei Valori at Intesa Sanpaolo and Giacomo Ranieri for valuable insights on classical methodologies for cash demand forecasting.

\bibliographystyle{IEEEtran}
\bibliography{biblio.bib}

@article{yasuda2023quantum,
  title={Quantum reservoir computing with repeated measurements on superconducting devices},
  author={Yasuda, Toshiki and Suzuki, Yudai and Kubota, Tomoyuki and Nakajima, Kohei and Gao, Qi and Zhang, Wenlong and Shimono, Satoshi and Nurdin, Hendra I and Yamamoto, Naoki},
  journal={arXiv preprint arXiv:2310.06706},
  year={2023}
}

@article{domingo2023taking,
  title={Taking advantage of noise in quantum reservoir computing},
  author={Domingo, Laia and Carlo, G and Borondo, Florentino},
  journal={Scientific Reports},
  volume={13},
  number={1},
  pages={8790},
  year={2023},
  publisher={Nature Publishing Group UK London}
}

@inproceedings{guo2024quantum,
  title={Quantum circuit ansatz: patterns of abstraction and reuse of quantum algorithm design},
  author={Guo, Xiaoyu and Muta, Takahiro and Zhao, Jianjun},
  booktitle={2024 IEEE International Conference on Quantum Software (QSW)},
  pages={69--80},
  year={2024},
  organization={IEEE}
}

@article{kornjavca2024large,
  title={Large-scale quantum reservoir learning with an analog quantum computer},
  author={Kornja{\v{c}}a, Milan and Hu, Hong-Ye and Zhao, Chen and Wurtz, Jonathan and Weinberg, Phillip and Hamdan, Majd and Zhdanov, Andrii and Cantu, Sergio H and Zhou, Hengyun and Bravo, Rodrigo Araiza and others},
  journal={arXiv preprint arXiv:2407.02553},
  year={2024}
}

@article{vitali2025quantum,
  title={Quantum Reservoir Computing for Credit Card Default Prediction on a Neutral Atom Platform},
  author={Vitali, Giacomo and Vercellino, Chiara and Viviani, Paolo and Terzo, Olivier and Montrucchio, Bartolomeo and Zaffaroni, Valeria and Cibrario, Francesca and Mattia, Christian and Ranieri, Giacomo and Sabatino, Alessandro and others},
  journal={arXiv preprint arXiv:2510.04747},
  year={2025}
}

@article{li2025quantum,
  title={Quantum reservoir computing for realized volatility forecasting},
  author={Li, Qingyu and Mukhopadhyay, Chiranjib and Bayat, Abolfazl and Habibnia, Ali},
  journal={arXiv preprint arXiv:2505.13933},
  year={2025}
}

@misc{iqm_spark,
  title        = {IQM Spark: Quantum Computing Platform},
  author       = {{IQM Quantum Computers}},
  year         = {2024},
  url          = {https://www.meetiqm.com/iqm-spark/},
  note         = {Accessed: 2026-03-31}
}

@article{taylor2018forecasting,
  title={Forecasting at scale},
  author={Taylor, Sean J and Letham, Benjamin},
  journal={The American Statistician},
  volume={72},
  number={1},
  pages={37--45},
  year={2018},
  publisher={Taylor \& Francis}
}

@article{chen2020temporal,
  title={Temporal information processing on noisy quantum computers},
  author={Chen, Jiayin and Nurdin, Hendra I and Yamamoto, Naoki},
  journal={Physical Review Applied},
  volume={14},
  number={2},
  pages={024065},
  year={2020},
  publisher={APS}
}

@article{rizzi2008simulation,
  title={Simulation of time evolution with multiscale entanglement renormalization ansatz},
  author={Rizzi, Matteo and Montangero, Simone and Vidal, Guifre},
  journal={Physical Review A—Atomic, Molecular, and Optical Physics},
  volume={77},
  number={5},
  pages={052328},
  year={2008},
  publisher={APS}
}

@article{hao2019machine,
  title={Machine learning made easy: a review of scikit-learn package in python programming language},
  author={Hao, Jiangang and Ho, Tin Kam},
  journal={Journal of educational and behavioral statistics},
  volume={44},
  number={3},
  pages={348--361},
  year={2019},
  publisher={SAGE Publications Sage CA: Los Angeles, CA}
}

@article{pedregosa2011scikit,
  title={Scikit-learn: Machine learning in Python},
  author={Pedregosa, Fabian and Varoquaux, Ga{\"e}l and Gramfort, Alexandre and Michel, Vincent and Thirion, Bertrand and Grisel, Olivier and Blondel, Mathieu and Prettenhofer, Peter and Weiss, Ron and Dubourg, Vincent and others},
  journal={the Journal of machine Learning research},
  volume={12},
  pages={2825--2830},
  year={2011},
  publisher={JMLR. org}
}

@article{miao2024quantumreservoirpy,
  title={Quantumreservoirpy: A software package for time series prediction},
  author={Miao, Stanley and Kulseng, Ola Tangen and Stasik, Alexander and Fuchs, Franz G},
  journal={arXiv preprint arXiv:2401.10683},
  year={2024}
}

@article{jaeger2007special,
  title={Special issue on echo state networks and liquid state machines.},
  author={Jaeger, Herbert and Maass, Wolfgang and Principe, Jose},
  year={2007},
  publisher={Elsevier Science}
}

@article{otieno2026quantum,
  title={A Quantum Reservoir Computing Approach to Quantum Stock Price Forecasting in Quantum-Invested Markets},
  author={Otieno, Wendy and Zagoskin, Alexandre and Balanov, Alexander G and Gongora, Juan Totero and Savel'ev, Sergey E},
  journal={arXiv preprint arXiv:2602.13094},
  year={2026}
}

@misc{qiskit_ugate,
  author       = {{IBM Quantum}},
  title        = {Qiskit UGate API Reference},
  year         = {2026},
  url          = {https://quantum.cloud.ibm.com/docs/en/api/qiskit/qiskit.circuit.library.UGate},
  note         = {Accessed: 2026-03-31}
}

@article{JMLR:v21:20-091,
  author  = {Romain Tavenard and Johann Faouzi and Gilles Vandewiele and
             Felix Divo and Guillaume Androz and Chester Holtz and
             Marie Payne and Roman Yurchak and Marc Ru{\ss}wurm and
             Kushal Kolar and Eli Woods},
  title   = {Tslearn, A Machine Learning Toolkit for Time Series Data},
  journal = {Journal of Machine Learning Research},
  year    = {2020},
  volume  = {21},
  number  = {118},
  pages   = {1-6},
  url     = {http://jmlr.org/papers/v21/20-091.html}
}

@article{senin2008dynamic,
  title={Dynamic time warping algorithm review},
  author={Senin, Pavel},
  journal={Information and Computer Science Department University of Hawaii at Manoa Honolulu, USA},
  volume={855},
  number={1-23},
  pages={40},
  year={2008}
}

@article{fujii2017harnessing,
  title={Harnessing disordered-ensemble quantum dynamics for machine learning},
  author={Fujii, Keisuke and Nakajima, Kohei},
  journal={Physical Review Applied},
  volume={8},
  number={2},
  pages={024030},
  year={2017},
  publisher={APS}
}

@article{nakajima2019boosting,
  title={Boosting computational power through spatial multiplexing in quantum reservoir computing},
  author={Nakajima, Kohei and Fujii, Keisuke and Negoro, Makoto and Mitarai, Kosuke and Kitagawa, Masahiro},
  journal={Physical Review Applied},
  volume={11},
  number={3},
  pages={034021},
  year={2019},
  publisher={APS}
}

@article{jaeger2001echo,
  title={The “echo state” approach to analysing and training recurrent neural networks-with an erratum note},
  author={Jaeger, Herbert},
  journal={Bonn, Germany: German national research center for information technology gmd technical report},
  volume={148},
  number={34},
  pages={13},
  year={2001},
  publisher={Bonn}
}

@article{maass2002real,
  title={Real-time computing without stable states: A new framework for neural computation based on perturbations},
  author={Maass, Wolfgang and Natschl{\"a}ger, Thomas and Markram, Henry},
  journal={Neural computation},
  volume={14},
  number={11},
  pages={2531--2560},
  year={2002},
  publisher={MIT Press}
}

@book{box2015time,
  title={Time series analysis: forecasting and control},
  author={Box, George EP and Jenkins, Gwilym M and Reinsel, Gregory C and Ljung, Greta M},
  year={2015},
  publisher={John Wiley \& Sons}
}

@book{hyndman2018forecasting,
  title={Forecasting: principles and practice},
  author={Hyndman, Rob J and Athanasopoulos, George},
  year={2018},
  publisher={OTexts}
}

@article{hochreiter1997long,
  title={Long short-term memory},
  author={Hochreiter, Sepp and Schmidhuber, J{\"u}rgen},
  journal={Neural computation},
  volume={9},
  number={8},
  pages={1735--1780},
  year={1997},
  publisher={MIT press}
}

@article{friedman2001greedy,
  title={Greedy function approximation: a gradient boosting machine},
  author={Friedman, Jerome H},
  journal={Annals of statistics},
  pages={1189--1232},
  year={2001},
  publisher={JSTOR}
}

@article{lukovsevivcius2009reservoir,
  title={Reservoir computing approaches to recurrent neural network training},
  author={Luko{\v{s}}evi{\v{c}}ius, Mantas and Jaeger, Herbert},
  journal={Computer science review},
  volume={3},
  number={3},
  pages={127--149},
  year={2009},
  publisher={Elsevier}
}

@misc{qiskit_aer,
  author       = {{Qiskit Development Team}},
  title        = {Qiskit Aer: High Performance Quantum Circuit Simulators},
  year         = {2024},
  url          = {https://qiskit.org/ecosystem/aer/},
  note         = {Accessed: 2026-03-31}
}

@misc{iqm_adonis,
  author       = {{IQM Quantum Computers}},
  title        = {IQM Adonis: Superconducting Quantum Processor},
  year         = {2025},
  url          = {https://iqm-finland.github.io/cirq-on-iqm/api/iqm.cirq\_iqm.devices.adonis.html},
  note         = {Accessed: 2026-03-31}
}

@article{iqm_noise_modeling,
  title   = {Noise-Robust Estimation of Quantum Observables in Noisy Hardware},
  author  = {Hosseinkhani, Amin and others},
  journal = {arXiv preprint arXiv:2503.06695},
  year    = {2025}
}

\end{document}